\documentclass{PoS}

\title{The $U_A(1)$ anomaly in high temperature QCD with chiral fermions on the lattice}

\ShortTitle{The $U_A(1)$ anomaly in high temperature QCD with chiral fermions on the lattice}

\author{\speaker{Sayantan Sharma}\\
        Brookhaven National Laboratory, \\
        Upton, New York 11973\\
        E-mail: \email{sayantans@quark.phy.bnl.gov}}

\author{Viktor Dick\\
        Bielefeld University\\
        Universit\"{a}tstasse 25, D33615 Bielefeld\\
        E-mail: \email{viktor@physik.uni-bielefeld.de}}

\author{Frithjof Karsch\\
        Brookhaven National Laboratory, Upton, New York 11973\\
        and Bielefeld  University, Universit\"{a}tstasse 25, D33615 Bielefeld\\
        E-mail: \email{karsch@bnl.gov}}

\author{Edwin Laermann\\
        Bielefeld University\\
        Universit\"{a}tstasse 25, D33615 Bielefeld\\
        E-mail: \email{edwin@physik.uni-bielefeld.de}}

\author{Swagato Mukherjee\\
        Brookhaven National Laboratory\\
        Upton, New York 11973\\
        E-mail: \email{swagato@bnl.gov}}

\abstract{ The magnitude of the $U_A(1)$ symmetry breaking is expected to affect the nature of $N_f=2$ QCD chiral phase transition. The explicit breaking of chiral symmetry due to realistic light quark mass is small, so it is important to use chiral fermions on the lattice to understand the effect of $U_A(1)$ near the chiral crossover temperature, $T_c$. We report our latest results for the  eigenvalue spectrum of 2+1 flavour QCD 
with dynamical M\"{o}bius domain wall fermions at finite temperature probed using the overlap operator on $32^3\times 8$ lattice. We check how sensitive the low-lying eigenvalues are to the sea-light quark mass. We also present a comparison with the earlier independent results with domain wall fermions.}

\FullConference{The 33rd International Symposium on Lattice Field Theory\\
		14 -18 July 2015\\
		Kobe International Conference Center, Kobe, Japan*}

\begin{document}

\section{Introduction}
The symmetries of different phases of QCD tend to decide the order parameter characterizing a phase transition. However a remarkable observation was made that the $U_A(1)$ symmetry in QCD, though anomalous, could affect the nature of phase transition~\cite{pw}. The effect for $N_f=2$ is most interesting. Depending on whether $U_A(1)$ is effectively restored at the chiral transition temperature, the order of the phase transition and its universality class or both can change. All these arguments are based on perturbative renormalization group studies~\cite{pw,bpv} or bootstrap analysis~\cite{naka} of model quantum field theory with the same symmetries as QCD. The coefficient of the $U_A(1)$ breaking term is just a parameter in this model which can  only be estimated non-perturbatively. A careful lattice study is essential to have a comprehensive understanding of this phase transition. For QCD with physical $u$, $d$ quark mass, chiral symmetry is only mildly broken since $m_{u,d}\ll\Lambda_{QCD}$. Hence it is important to use fermions with exact chiral and flavor symmetries on the lattice to address the issue of the fate of $U_A(1)$ near the chiral crossover transition temperature $T_c$. We use domain wall fermions, which have nearly exact chiral symmetry on the lattice, to study this problem. To unambiguously identify the zero and the near-zero modes of the domain wall fermions we use the overlap fermion operator as a probe on the domain wall sea quarks. Our study suggests that the $U_A(1)$ is not effectively restored at $T_c$.
 
\section{What we have learnt from QCD with Highly Improved Staggered Quarks}
\begin{figure}[h]
\begin{center}
\includegraphics[scale=0.75]{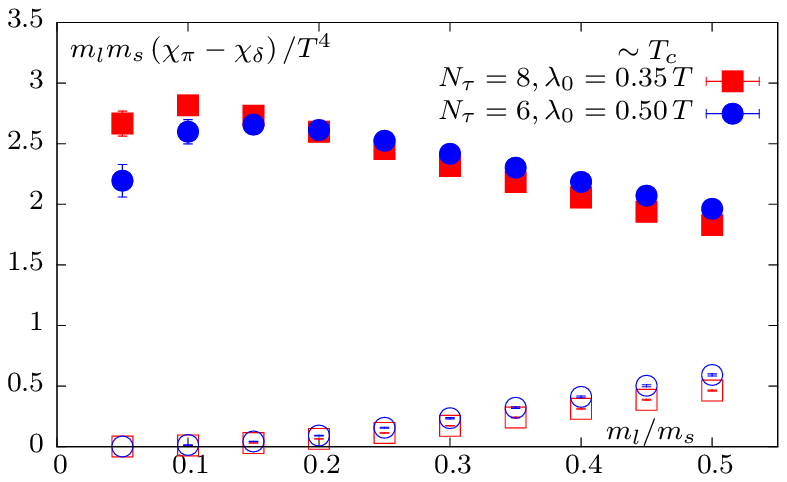}
\includegraphics[scale=0.75]{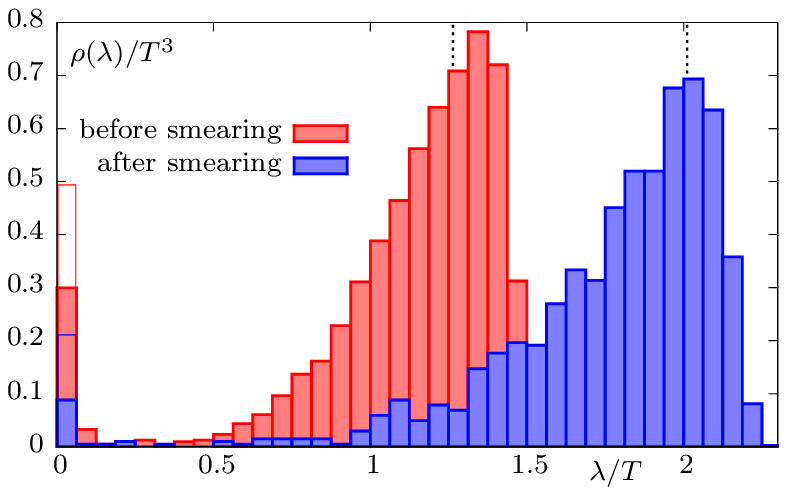}
\caption{The contribution of the infrared ($<\lambda_0$) Dirac modes to $U_A(1)$ breaking is dominant near $T_c$ (left panel). In the right panel, the eigenvalue spectrum of the HISQ configurations measured by the overlap fermions are shown on unsmeared gauge configurations (red) and after performing 10 levels of HYP smearing (blue) on them. These are taken from ~\cite{viktor}. }
\label{fig:hisq}
\end{center}
\end{figure}

 We recall here some of our key results from a similar study we did with $N_f=2+1$ QCD with Highly Improved Staggered Quarks (HISQ) from~\cite{viktor}. Since the relation between the zero modes of the HISQ operator to the underlying topology of the gauge fields is not explicit, we employed overlap fermions to study the eigenspectrum of the HISQ ensembles. We found that $U_A(1)$ remains broken until $\sim 1.5~ T_c$. The infrared modes of the Dirac operator mainly contributed to its breaking near $T_c$ since the contribution of the higher modes were small in the light quark mass region as is evident from the left panel of Fig. \ref{fig:hisq}. The infrared part of the spectrum has contributions both from the near-zero as well as from the infrared part of the bulk spectrum rising  as $|\lambda|$ near $T_c$ and it is not possible to disentangle their relative contributions. However the near-zero modes become sparse at $1.5~T_c$ and start separating from the bulk.  It is thus necessary to understand whether these near-zero modes are physical or lattice artifacts. In order to do so, we performed HYP smearing of the gauge configurations at $1.5 ~T_c$. Smearing is known to remove unphysical localized structures or dislocations present in the gauge configurations. The right panel of Figure~\ref{fig:hisq} shows the eigen density measured using the overlap operator on HISQ gauge configurations before and after smearing. Smearing eliminates some zero and near-zero modes. However it does not eliminate them completely, but reduces their 
density. The average radii of the instantons represented by the fermion zero modes do not change within errors. Smearing is known to eliminate the small instantons which may explain the difference in the spectrum before and after smearing. The near-zero modes are seen to be associated with an instanton-antiinstanton pair or can be motivated from a quasi-instanton picture~\cite{yama}. Our previous study~\cite{viktor} suggests that the average instanton density is $0.147(7) fm^{-4}$ at $1.5~ T_c $, hence the dilute instanton gas description is already appropriate for the QCD medium at $1.5~ T_c$. Similar features in eigenspectrum are also observed with stout-smeared staggered quarks at $T>T_c$ on even finer lattices~\cite{ivan}.

\vspace{-0.8cm}

\section{Why chiral fermions}
It is important to understand these results with respect to those obtained with exact chiral fermions on the lattice. The eigenvalues of the staggered fermions show four-fold degeneracy in the continuum limit and are believed to produce the correct $\eta'$ mass as shown explicitly in Schwinger model calculations~\cite{duerr}. However, on a finite lattice the near-zero part of the HISQ eigenspectrum could have more degeneracies which may render the continuum axial $U_A(1)$ explicitly broken.  We therefore study the eigenspectrum of $N_f=2+1$ M\"{o}bius domain wall fermion configurations with the overlap operator. The eigen densities of some of these ensembles with $m_\pi=200$ MeV have been measured and reported in~\cite{dw}. 
It is also evident from this work that $U_A(1)$ is not effectively restored at $T\gtrsim T_c$ and a peak of near-zero modes develops at $T \sim 1.1~T_c$, consistent with our observation with the HISQ fermions. However with the $5$ dimensional domain wall fermion operator, the identification of the near-zero modes is not unambiguous due to residual mass effects. It is thus important to identify them with the overlap operator. Moreover it is interesting to study the quark mass dependence of these results especially when one goes to the physical quark mass.

\section{Numerical details}
The configurations used in this work were generated with 
M\"{o}bius domain wall fermions with Iwasaki gauge action with dislocation suppressing determinant.
They have been taken from~\cite{dw}. 
The pion masses for these sets were $135$ and $200$ MeV respectively. The overlap operator was realized by calculating the sign function exactly with the eigenvalues of $D_W^\dagger D_W$ for low modes and representing the higher modes with a Zolotarev Rational function with $15$ coefficients. The sign function was measured to a precision of about $10^{-10}$ and the violation of the Ginsparg Wilson relation was of the same order of magnitude for each configuration. We chose the domain wall height appearing in the overlap operator $M=1.8$, which gave the best approximation to the sign function and satisfied the Ginsparg Wilson relation with the best precision. On each configuration $50$ eigenvalues of $D_{ov}^\dagger D_{ov}$ were measured using the Kalkreuter-Simma Ritz algorithm~\cite{ks}.
We used about $100-150$ configurations at each temperature and for each value of the pion mass.
 
\section{General features of the domain wall spectrum and $U_A(1)$ breaking}
\begin{figure}[h]
\begin{center}
\includegraphics[scale=0.5]{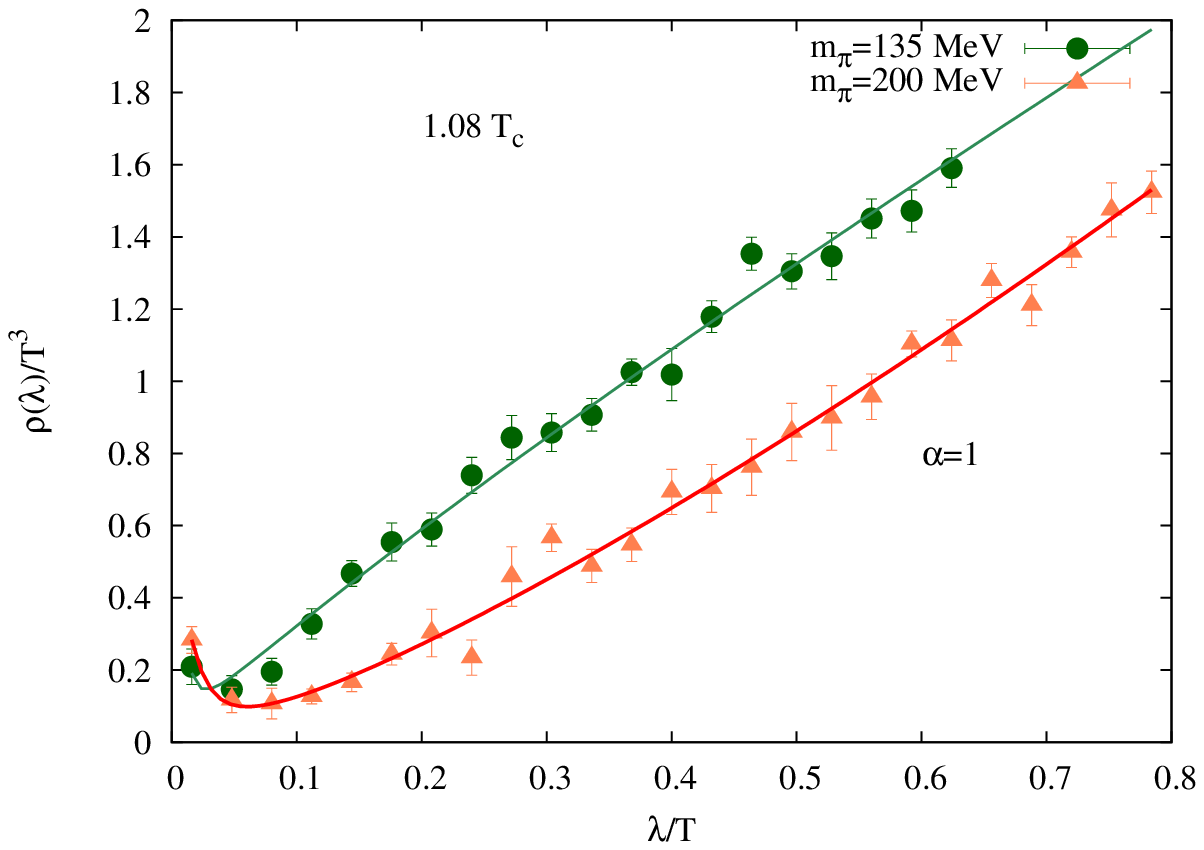}
\includegraphics[scale=0.5]{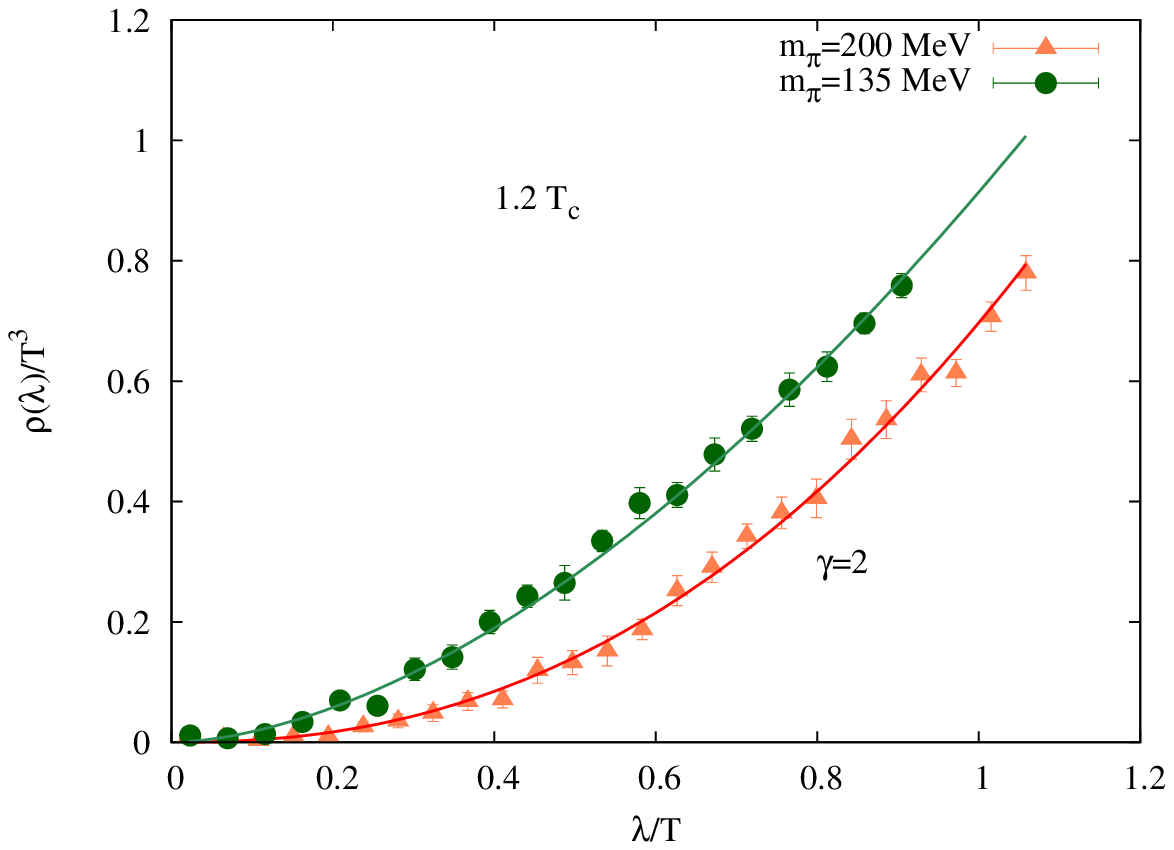}
\caption{The eigenvalue distribution of  M\"{o}bius domain wall fermions at $1.08~T_c$ (left panel) and $1.2~ T_c$ (right panel) for two different sea-quark masses measured with the overlap operator. }
\label{fig:eigdw}
\end{center}
\end{figure}

The overlap eigen density measured on the domain wall fermion ensembles at $1.08 ~T_c$ and $1.2~ T_c$ for two different pion masses are shown in Figure \ref{fig:eigdw}. The general features are the presence of near-zero (non-analytic) modes and the bulk which is analytic in $\lambda$.
We fit the eigenspectrum with the ansatz $\rho(\lambda)=\frac{A}{\lambda^2+\rho^2}+c |\lambda|^\gamma$. 
The fit results for $\gamma$ which characterize the leading order rise of the bulk modes, have a significant temperature dependence. At $1.08 ~T_c$, the value is $\gamma\sim 1$, which changes to $\gamma\sim 2$ at $1.2~ T_c$ consistent with our earlier observation from the HISQ eigenspectrum measured with the overlap operator~\cite{viktor}. Moreover $\gamma$ is insensitive to quark mass effects so its value would not change in the chiral limit as well. The analytic part of the eigenvalue spectrum of the QCD Dirac operator at finite temperature has been analytically studied only recently~\cite{aoki}. In the phase of restored chiral symmetry, using chiral Ward identities on $3$-point correlation functions, it has been derived that the leading order analytic part of the eigen density goes as $|\lambda|^3$. It was further shown that the $U_A(1)$ breaking effects are invisible in the scalar and pseudo-scalar sector in up to $6$-point correlation functions~\cite{aoki} with this analytic dependence. From our results it is evident that the $U_A(1)$ breaking effects from the analytic part of the eigenvalue spectrum survive even at $1.2 ~T_c$. 
\begin{figure}[h]
\begin{center}
\includegraphics[scale=0.5]{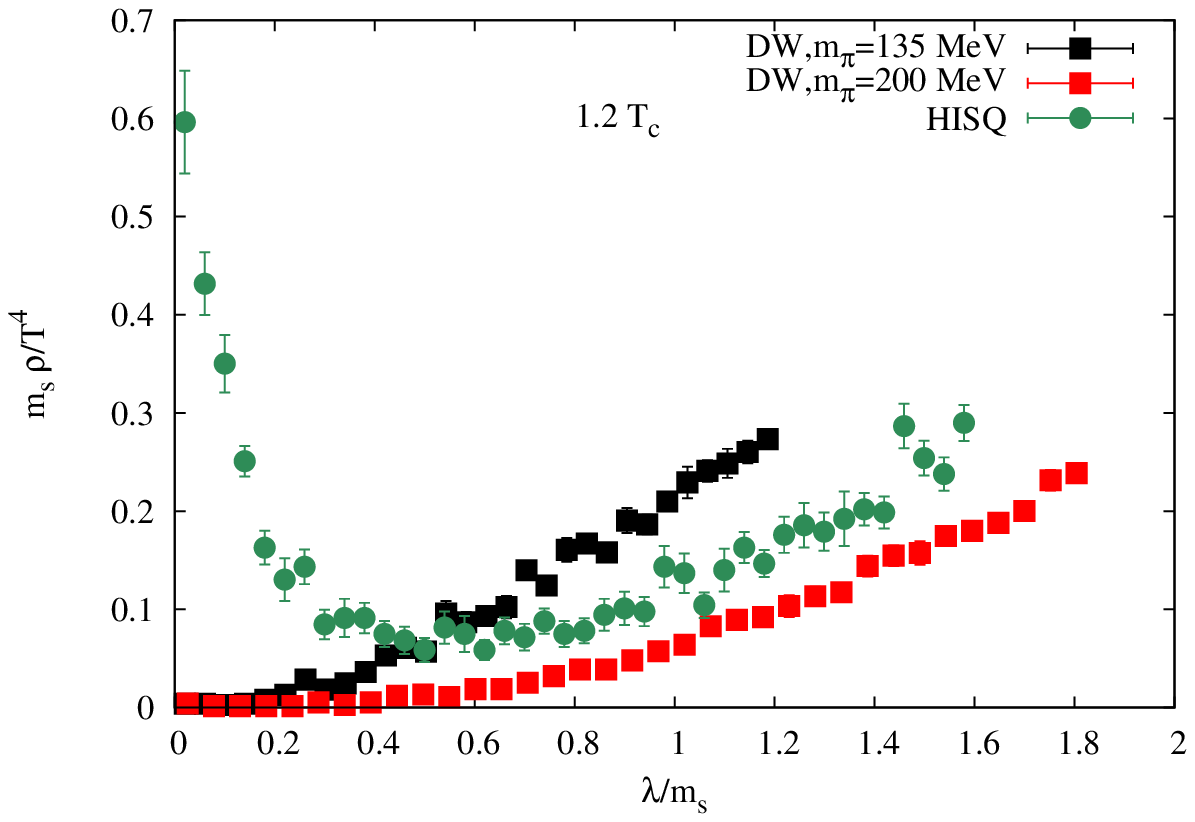}
\includegraphics[scale=0.5]{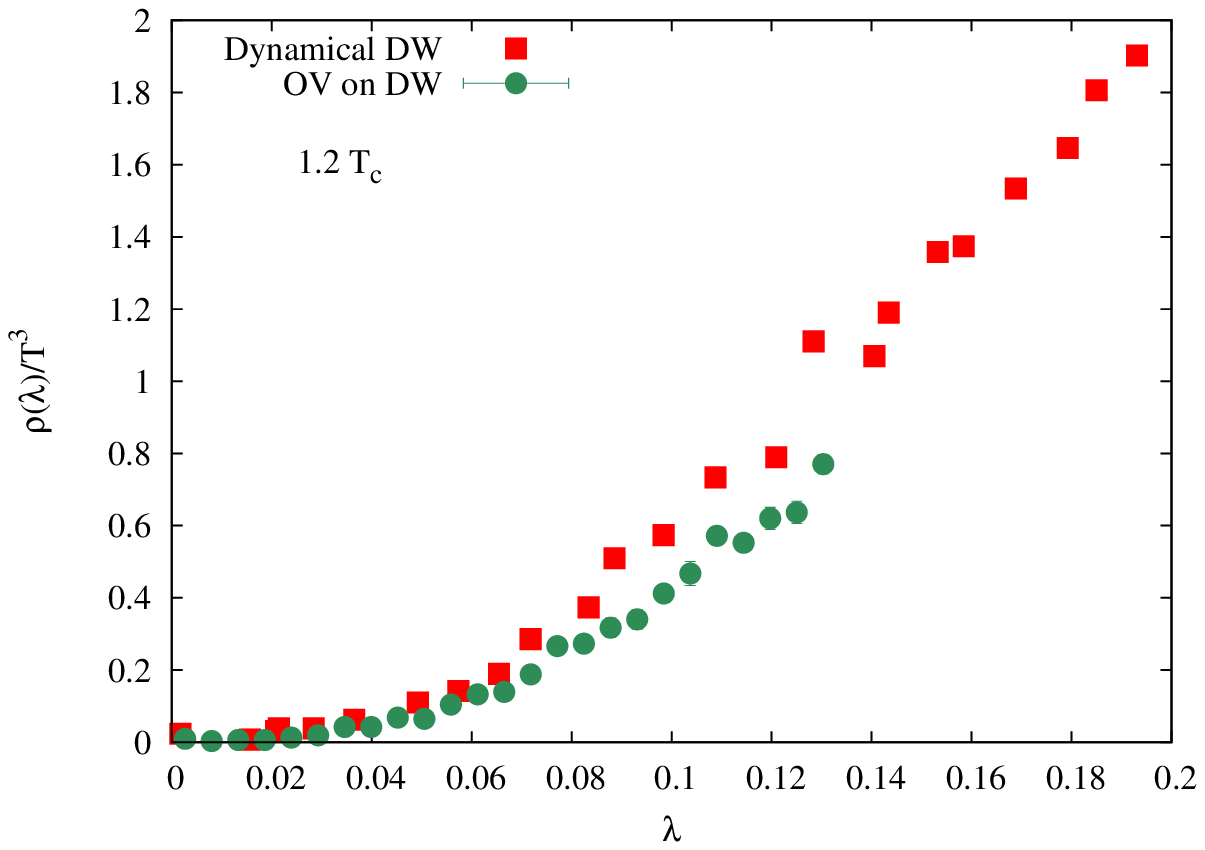}
\caption{The renormalized eigenvalue distribution of the overlap on domain wall fermion configurations compared to overlap on HISQ configurations (left panel) and dynamical domain wall fermion configurations from ~\cite{dw}(right panel) at $1.2~ T_c$. }
\label{fig:dwcomp}
\end{center}
\end{figure}
We next compare the eigenvalue spectrum of the domain wall fermion operator and the overlap operator on the same configurations at $m_\pi=200$ MeV in Figure \ref{fig:dwcomp}. There is a fairly good  agreement between the eigenspectra implying that the lattice artifacts when using the overlap operator are under control. We also compare the renormalized eigenspectrum of the domain wall and the HISQ fermions measured by us using the overlap operator~\cite{viktor} in the right panel of Figure \ref{fig:dwcomp}.  The eigenvalues are renormalized by the strange quark mass $m_s$, the determination of which is explained in the next paragraph. The bulk modes for the domain wall fermion configurations with $m_\pi=200$ MeV rather than $m_\pi=135$ MeV agree well with those of the HISQ configurations with a root mean square $m_\pi=160$ MeV. This is not very surprising; the effective quark masses in the HISQ ensembles could be larger than the rms mass. These results give us confidence that the bulk spectrum measured with overlap fermions on different sea-quark ensembles are similar, with the exponent $\gamma$ characterizing the leading-order behavior not being very sensitive to lattice cut-off effects.  

\begin{figure}[h]
\begin{center}
\includegraphics[scale=0.55]{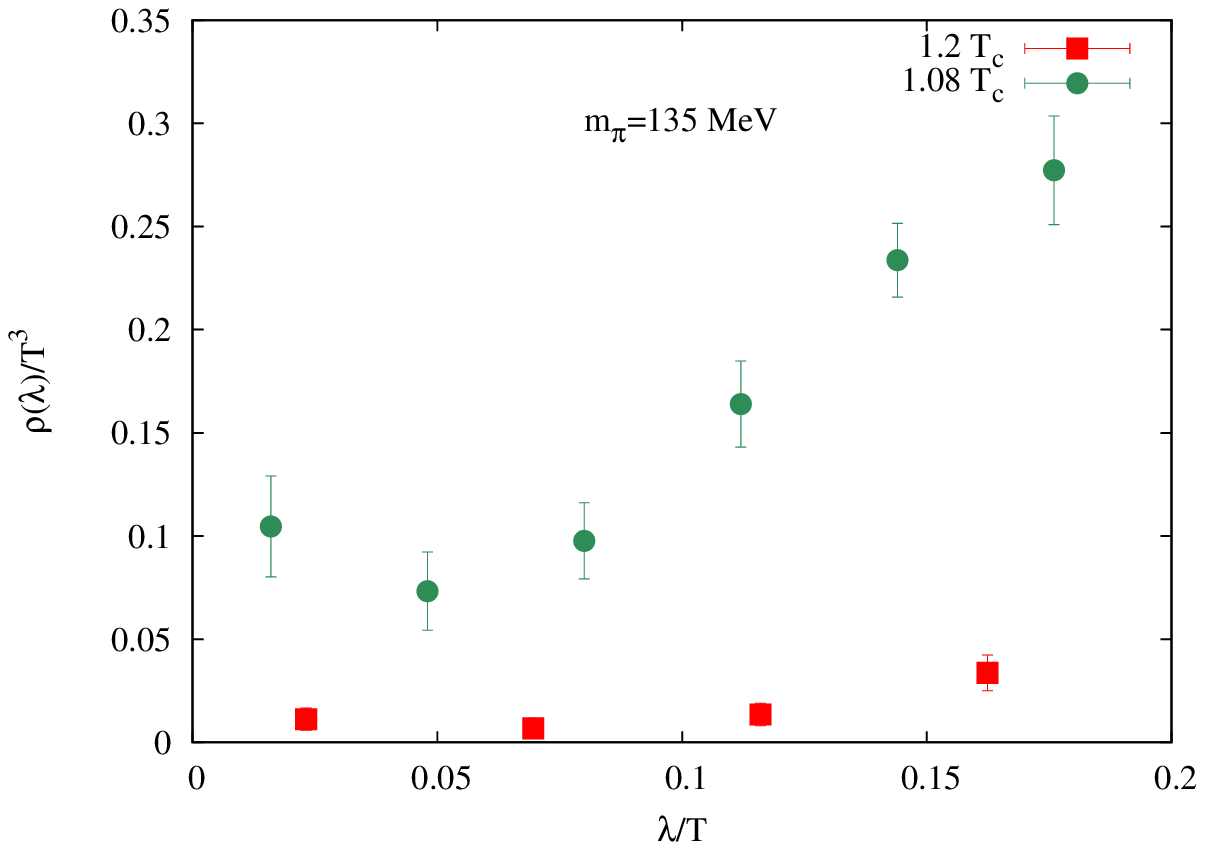}
\includegraphics[scale=0.55]{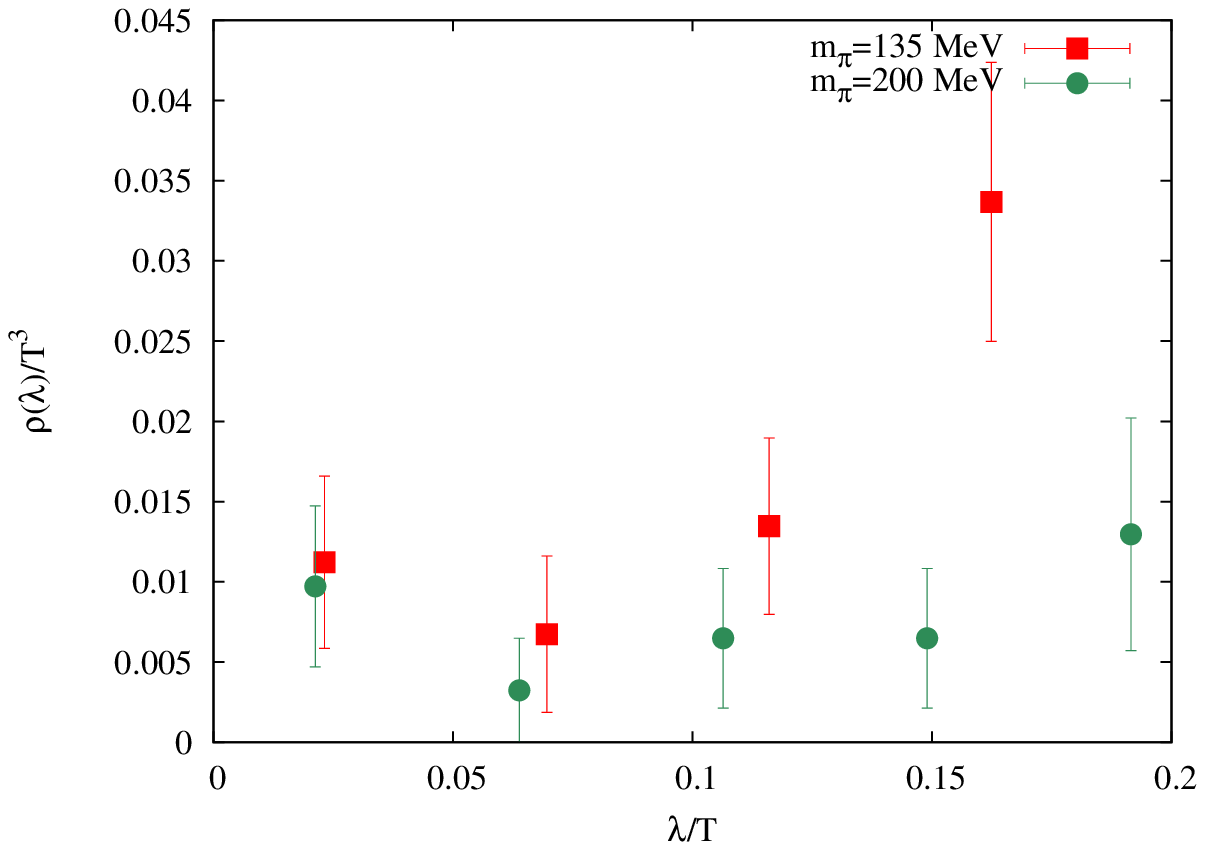}
\caption{The near-zero eigenvalue distribution for physical quark mass at two different temperatures (left panel) and as a function of sea quark mass at $1.2 ~T_c$ (right panel).}
\label{fig:nearzdw}
\end{center}
\end{figure}

\begin{figure}[h]
\begin{center}
\includegraphics[scale=0.5]{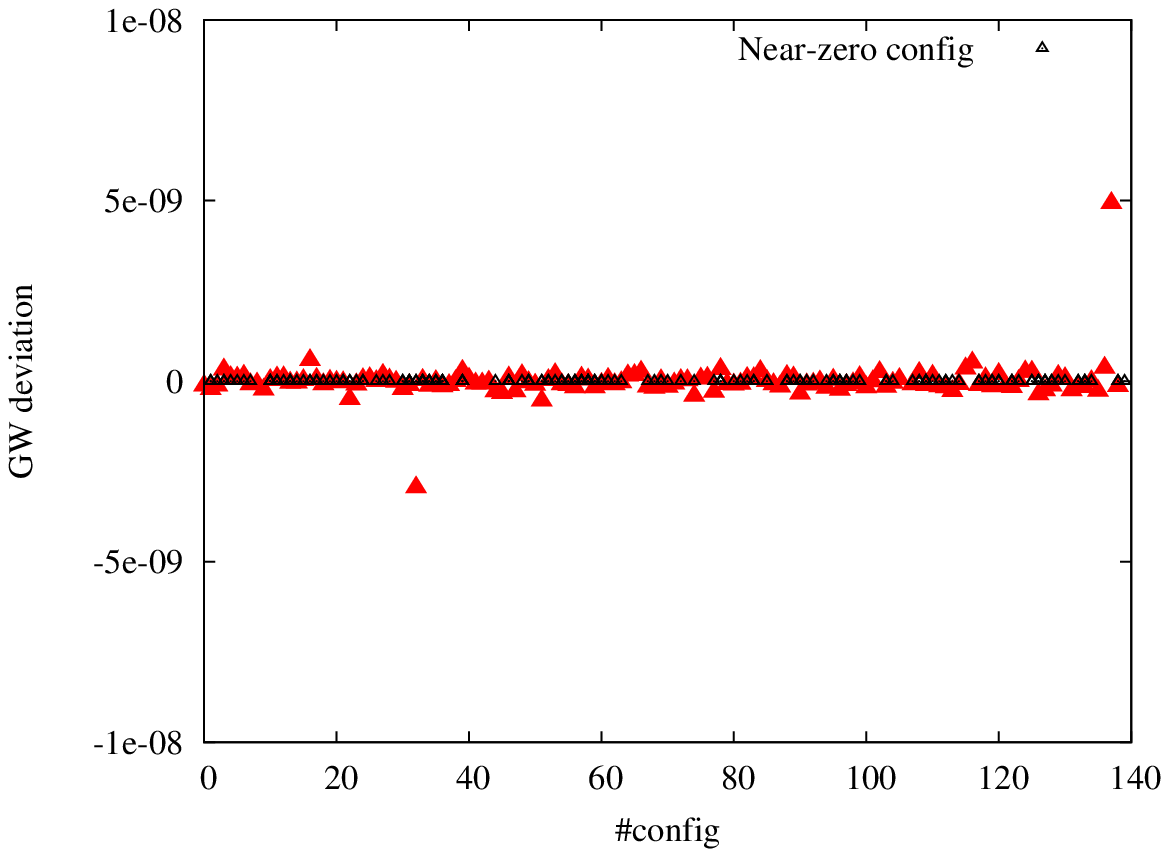}
\includegraphics[scale=0.5]{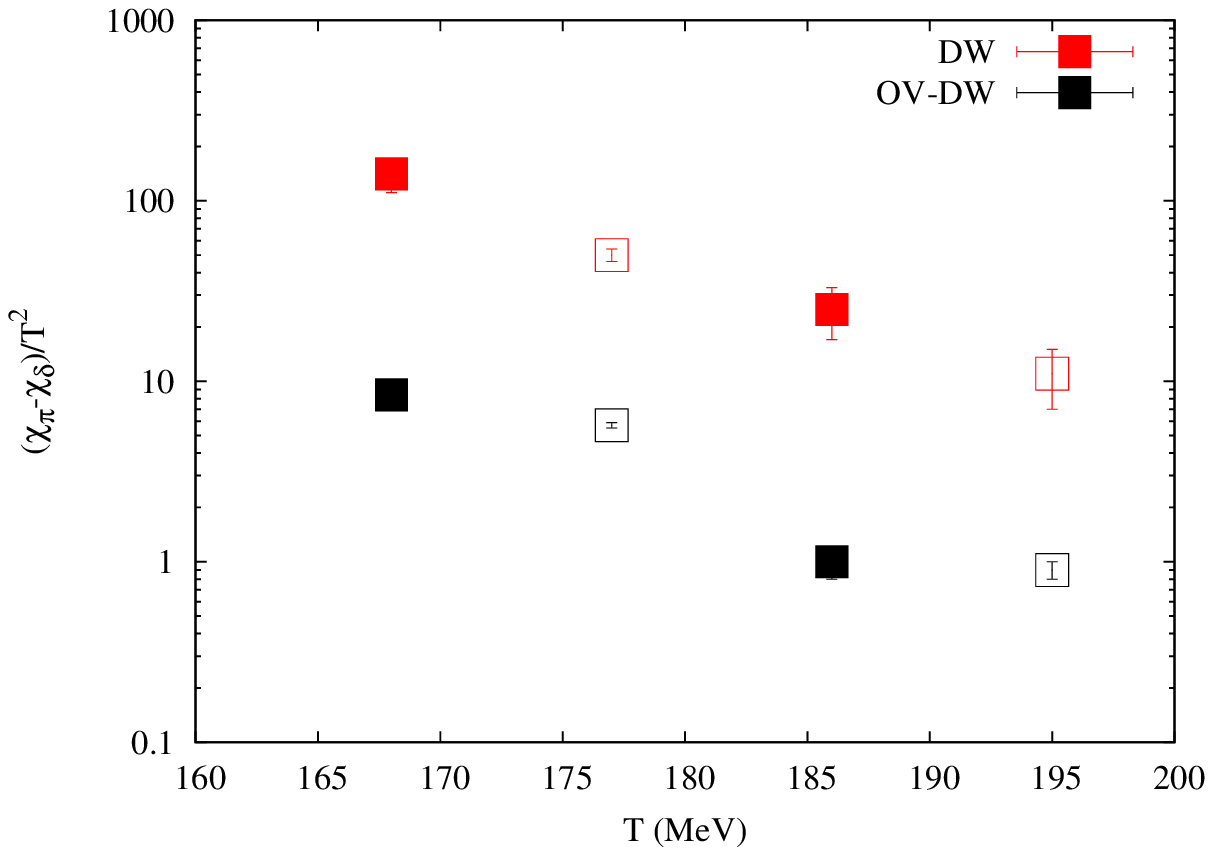}
\caption{The red points show the violation of Ginsparg Wilson relation for the overlap operator for each domain wall fermion configuration 
at $1.08 ~T_c$ for physical quark mass (left panel) and the black points mark those configurations which have 
near-zero modes. In the right panel the temperature dependence of $(\chi_\pi-\chi_\delta)/T^2$ is shown for $m_l/m_s\sim 1/12$ (open symbols) 
and $m_l/m_s\sim 1/27$ (solid). }
\label{fig:gwdw}
\end{center}
\end{figure}

\begin{figure}[h]
\begin{center}
\includegraphics[scale=0.4]{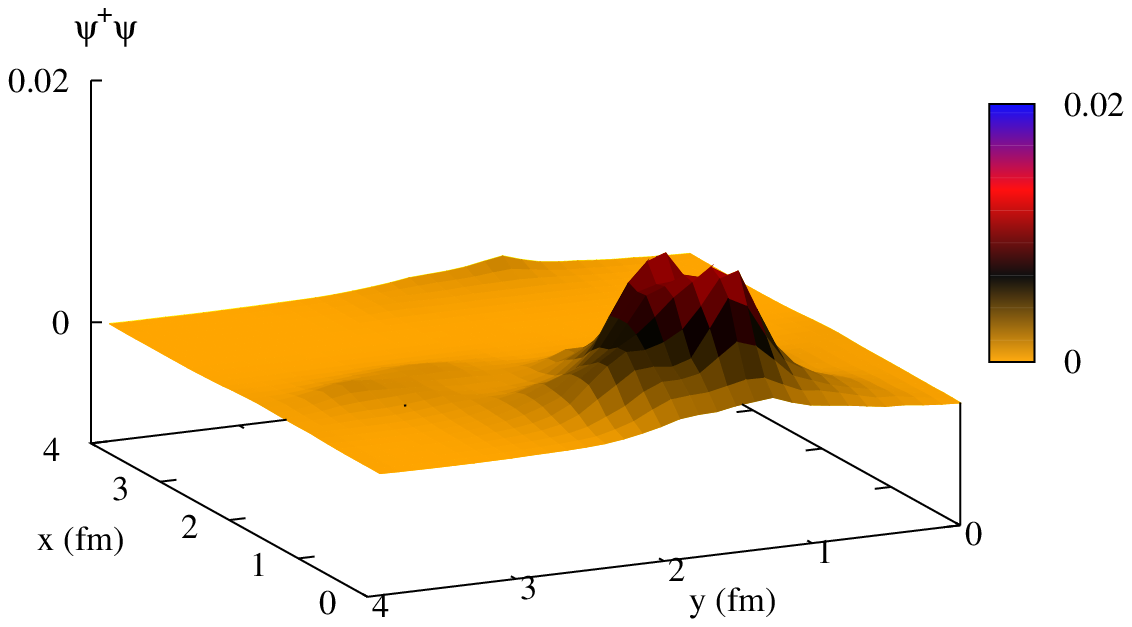}
\includegraphics[scale=0.4]{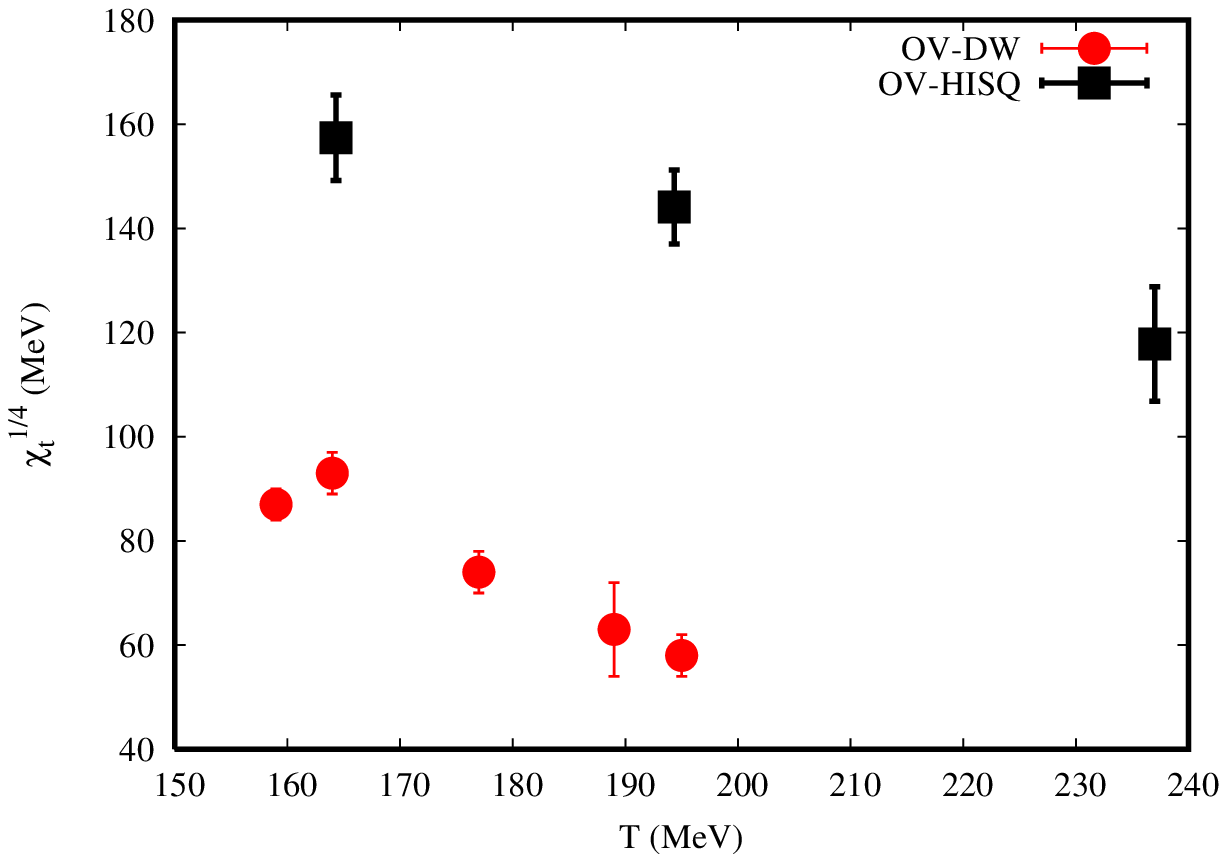}
\caption{The typical profile for a near-zero mode at $1.2 ~T_c$ (left panel). The topological susceptibility as a function of temperature compared between HISQ (from \cite{viktor}) and domain wall fermions (right panel) measured with the overlap operator. }
\label{fig:dwprof}
\end{center}
\end{figure}

We look more carefully at the near-zero modes since it is evident from the right panel of Figure \ref{fig:dwcomp} and from ~\cite{cossu}, that it is sensitive to lattice cut-off effects of the valence quarks. For the domain wall fermions, the near-zero modes are suppressed already at $1.2~ T_c$ compared to the HISQ fermions. The left panel of Figure \ref{fig:nearzdw} shows the near-zero peak as a function of temperature for $m_\pi=135~$ MeV. The peak height reduces by a third, as the temperature changes from $1.08 ~T_c$ to $1.2 ~T_c$. The right panel shows the peak height dependence on the sea quark mass at $1.2~T_c$. The near zero peak shows little sensitivity to the sea-quark mass when the light quark mass, $m_l$ changes from $\sim m_s/12$ to $\sim m_s/27$.    
It has been earlier observed on small volume M\"{o}bius domain wall configurations, that the near-zero modes of the valence overlap modes could arise due to violations of the Ginsparg Wilson relation of the overlap operator~\cite{cossu}. We therefore compare the Ginsparg Wilson relation violation in those configurations which have near-zero modes marked by black triangles in the left panel of Figure \ref{fig:gwdw} to the average violation observed at $1.2 ~T_c$. We do not observe any strong correlation between the violation of Ginsparg Wilson relation and the occurrence of near-zero modes. 
We next study the $U_A(1)$ violation as a function of $T$ in the domain wall fermion ensembles. The degeneracy of the pion and delta 
correlation functions is a possible signature for the effective restoration of $U_A(1)$. The difference of the integrated correlators 
of pion and delta mesons, $\chi_\pi-\chi_\delta$ normalized by $T^2$, determined from the eigenvalues of the overlap operator are plotted 
for different light quark masses, $m_l$ in the right panel of Figure \ref{fig:gwdw}. These values are compared with the values obtained 
on the same configurations from inversion of the domain wall operator~\cite{dw}. To make a faithful comparison, we have tuned the overlap 
valence quark mass $m_l$ by comparing the RG invariant quantity $\frac{m_s\langle\bar\psi\psi\rangle_l-m_l\langle\bar\psi\psi\rangle_s}{T^4}$ 
keeping $m_s/m_l\sim 27,12$ calculated with the overlap eigenvalues to those obtained in ~\cite{dw} by inverting the domain wall operator. 
The consistency between them suggests that the values we get from the overlap eigenvalues are not mere lattice artifacts. The fermion 
near-zero modes have some sort of a two peak structure with equal and opposite chiralities shown in the left panel of Figure 
\ref{fig:dwprof}. However the two peaks are not well separated. If indeed these arise due to overlap between localized fermion 
zero-modes associated with a closely located instanton-antiinstanton pair, their structure suggests that the dilute gas picture 
of such topological objects only begins to set in at $T> 1.2 ~T_c$. This is in agreement with our studies with HISQ fermions. 
We also show our results of the topological susceptibility $\chi_t$ measured on the domain wall fermion configurations and compare 
with our earlier HISQ measurements in the right panel of Figure \ref{fig:dwprof}. $\chi_t$ is comparatively smaller for the domain 
wall fermions but for both the temperature dependence for $T\leq 1.2~ T_c$  is still not compatible  with expectations from a dilute 
gas of instantons.

\section{Conclusions and outlook}
In this work we have studied in detail, the eigenvalue spectrum of QCD with two light quark flavours with nearly chiral M\"{o}bius domain wall fermions, to check whether $U_A(1)$ is effectively restored near $T_c$. Using overlap operator, we could unambiguously distinguish the zero from the near-zero modes of the domain wall fermions. We find that $U_A(1)$ is not effectively at $T\gtrsim T_c$ even when the light quark mass changes from $m_s/12$ to $m_s/27$. 
 $U_A(1)$ is broken due to the small eigenvalues which have contributions from the analytic and non-analytic part of eigenvalue spectrum. While the (analytic)contribution from the bulk part is very robust, i.e. insensitive to the quark mass and lattice discretization effects, one needs to further study the sensitivity of the near-zero eigenvalues to lattice cut-off effects and volume.

\section{Acknowledgments} 
This work has been supported in part through contract DE-SC0012704  with the U.S. Department of Energy, the BMBF under grants  05P12PBCTA and 05P15PBCAA. Numerical calculations have been performed using the GPU cluster at Bielefeld University. The GPU codes used in our work were in part based on some publicly available QUDA libraries~\cite{quda}.


\begin{thebibliography}{99}
\bibitem{pw} 
R.\ D.\ Pisarski and F.\ Wilczek,  Phys.\ Rev.\  D 29, 338 (1984).
\bibitem{bpv}
A.\ Butti, A.\ Pelissetto, E.\ Vicari, JHEP 0308, 029 (2003); A.\ Pelissetto, E.\ Vicari, Phys. Rev. D 88, 105018 (2013); M.\ Grahl and D.\ H.\ Rischke, 
Phys.\ Rev.\ D 88,  056014 (2013); T.\ Sato and N.\ Yamada, Phys.\ Rev.\ D 91, 034025 (2015).
\bibitem{naka}
Y.\ Nakayama and T.\ Ohtsuki, Phys.\ Rev.\ D 91, 021901 (2015).
\bibitem{viktor}
V.\ Dick et. al., Phys. Rev. D 91, 094504 (2015).
\bibitem{yama}
T.\ Kanazawa and N.\ Yamamoto, Phys. Rev. D 91,  105015 (2015); arxiv:1508.02416. 
\bibitem{ivan}
A.\ Alexandru and I.\ Horvath, Phys. Rev. D 92, 045038 (2015).
\bibitem{duerr}
S. D\"{u}rr, Phys.Rev. D85  114503 (2012).
\bibitem{dw}
M. I. Buchoff et. al., Phys. Rev. D 89, 054514 (2014); T.\ Bhattacharya et. al.,  Phys. Rev. Lett. 113, 082001 (2014).
\bibitem{ks}
T.\ Kalkreuter and H.\ Simma, Comput. Phys. Commun. 93, 33 (1996).
\bibitem{aoki}
  S.\ Aoki, H.\ Fukaya, Y.\ Taniguchi, Phys.\ Rev.\ D 86, 114512 (2012).
\bibitem{cossu}
G. Cossu et. al., PoS LATTICE2014 (2015) 210, [arxiv:1412.5703 [hep-lat]]; A. Tomiya et. al., PoS LATTICE2014 (2015) 211, [arXiv:1412.7306 [hep-lat]]; Proceedings of this Conference
\bibitem{quda}
 M.\ A.\ Clark et. al., Comput.\ Phys.\ Commun.\ 181, 1517 (2010).
\end{thebibliography}
\end{document}